\documentstyle[11pt,epsf]{article}
\def\ll{\label}
\def\re{\ref}
\def\c{\cite}
\def\b{\begin}

\def\r1{(\ref{$1})}
\def\ot{\otimes}

\def\kap{\kappa}

\def\ti{\tilde}

\def\ep{\epsilon}
\def\th{\theta}
\def\ba{\begin{array}{c}}
\def\e{\end}
\def\sk{\smallskip}
\def\ea{\end{array}}
\def\pr{\prod}
\def\ni{\noindent}
\def\si{\sigma}
\def\da{\dagger}
\def\De{\Delta}
\def\de{\delta}
\def\bet{\beta}
\def\ov{\over}
\def\ha{{1\over 2}}

\def\l{\left}
\def\l({\left(}
\def\r){\right)}
\def\r{\right}
\def\rw{\rightarrow}

\def\la{\lambda}
\def\al{\alpha}

\def\be{\begin{equation}}
\def\bc{\begin{center}}
\def\ec{\end{center}}
\def\bit{\begin{itemize}}
\def\eit{\end{itemize}}
\def\ee{\end{equation}}
\def\ed{\end{document}}
\def\bea{\begin{eqnarray}}
\def\eea{\end{eqnarray}}
\def\efr{\end{flushright}}

\begin{document}
\title{ Unifying structures in  quantum integrable systems
} 

\author{
Anjan Kundu \footnote {email: anjan@tnp.saha.ernet.in} \\  
  Saha Institute of Nuclear Physics,  
 Theory Group \\
 1/AF Bidhan Nagar, Calcutta 700 064, India.
 }
\maketitle
\vskip 1 cm

\rightline{SINP/TNP/16-97, solv-int/9710018}

\begin{abstract} 
Basic concepts of quantum integrable systems (QIS)
 are presented stressing on 
the  unifying  structures underlying such diverse models.  
Variety of ultralocal and nonultralocal  models is shown to be 
 described by a few basic 
 relations defining novel algebraic entries. 
Such   properties
 can generate and classify  integrable models systematically
 and also help to  solve exactly their eigenvalue problem 
in an almost model-independent way.
The unifying thread stretches also beyond the QIS to establish its deep 
connections with statistical models,  conformal field theory 
 etc.  as well as with 
abstract mathematical objects like quantum group, braided or quadratic
algebra.
%
\end{abstract}


\section{Introduction}
\setcounter{equation}{0}

A number of quantum models of diverse nature shows the important property
of integrability in low dimensions, enabling  us to solve  them  
exactly. Examples of such models  may extend 
from field models like sine-Gordon  (SG), nonlinear Schr\"odinger
 equation (NLS), derivative NLS (DNLS)
 etc. to the discrete models like Toda chain
(TC), relativistic TC, isotropic ($XXX$) 
or anisotropic spin chains ($XXZ, XYZ$) etc. Similarly
they may have varied basic commutation relations ranging from ultralocal to
nonultralocal models. However, the fascinating feature of such models, as
revealed in recent years, is their common
 underlying algebraic structures defined through a basic relation called
Yang-Baxter equation.  The unifying spirit goes also beyond the domain of
the quantum integrable systems (QIS)
 to  reveal its deep relations with variety of other disciplines, which were
 thought to be completely unrelated in the recent
past. Thus in one hand the QIS is deeply connected with abstract
mathematical objects like quantum group, noncocommutative Hopf algebras,
quadratic and braided algebra, universal ${\cal R}$-matrix etc. and on the
other hand it has intimate relation with
 stat-mech problems, conformal field theory (CFT), knot and braid theories
etc.

The starting of   QIS-age  
  should possibly  be counted from  1931, 
when Hans Bethe  was the first  to solve  exactly  
an  interacting many-body quantum problem, e.g. the isotropic
Heisenberg spin chain,
 through his by now celebrated  ansatz called the  Bethe ansatz \c {bethe}.
This was followed by many other applications and successes of the method 
in solving various physically interesting  models \c{qis}
like  $XXZ$-spin chain \c{xxz}, many-body problem with attractive 
\c{delta+} or repulsive  \c{delta-} $\de$-function interactions,
Hubbard model \c{hubbard}, Kondo problem  \c{kondo} etc.
Next came the  landmark discovery of Rodney Baxter,
in which he not only solved the completely anisotropic $XYZ$
 spin chain  but also
established its deep link with a stat-mech problem, e.g. the eight-vertex
model \c{xxz}.   
However, 
the  genuine  theory of  QIS based on all the preceding achievements
in quantum as well as classical integrable  theories
  was   started taking shape   
   only in the late seventies \c{skly79}\c{thacker}.
 Synthesising the concept of the
Lax-operator, borrowed from the classical theory, along with the Bethe ansatz
the quantum inverse scattering method (QISM) was developed mainly through 
 the works of the
Russian school, where the celebrated 
 Yang-Baxter equation  plays the central role and an algebraic 
formulation of the Bethe ansatz was invented for covering  wider
 range of models including field theoretic models  \c{fadrev}. 
The connection of the integrable system with the conformal field theory (CFT)
\c{zamcft} \c{vegacft} as well as with  the quantum group \c{drinfeld} 
and the braid and knot 
theory \c{wadatipr} was understood quite recently. 
Thanks to this  ever-growing progress, today the theory of QIS is considered 
to be a  major and 
important branch of theoretical and mathematical physics, 
with many research
groups all over the globe  engaged in  active research in the field.

 We  consider the 
   notion of  integrability
in the Liouville sense, i.e.
  call a system integrable, when  the number of 
  independent   
 conserved quantities  coincides with the degree
of freedom   of the system.
 In the quantum case they  correspond to 
  conserved operators with one being the  
  Hamiltonian, commuting with them.
Moreover,  their independence demands that all of them must  form a 
mutually commuting set of operators, which in  field models 
 are
infinite in number.   We will see that the quantum
Yang-Baxter equation, which plays the role of the basic equation
in QIS,  ensures in fact such a property.
Though in physics we are interested mainly in the Hamiltonian 
 and  its eigenvalue solutions,
  a completely integrable system  actually 
guarantees much more.  it  should in principal
  give   the whole hierarchy
of conserved operators
along with their  exact eigenvalue solutions, which is one of the
main aims of the QISM. 

We present here the basic concepts of QISM with a major
  stress on 
the  unifying scheme  for both  
 ultralocal and nonultralocal quantum models. We  elaborate 
on the fascinating and novel algebraic  structures revealed by the theory,
which help to generate and solve such models in a systematic way.  
We also focus on the intriguing 
connections of the theory with the stat-mech and CFT models. 
The  section headings  explain the topics dealt in each of them.

\section{Lax operator and examples of  integrable systems}
\setcounter{equation}{0}

The central idea  of classical ISM \c {soliton} 
is that, instead of dealing with  the 
nonlinear equation in $(1+1)$-dimensions directly, it constructs 
the corresponding linear
scattering problem
\be
{\cal T}_x(x,\lambda)= L(u(x,t),p(x,t), \lambda)~ {\cal T}(x,\lambda)
,\ll{laxeq} \ee
where  the Lax operator $L(u,p, \lambda)~$ depending on  the fields $u, p$  
and the { spectral parameter} $\la$ contains all  information
about the original nonlinear system and may serve therefore 
as the representative of a concrete model. The field $u$ 
acts as the scattering potential.
The aim of  ISM is to solve the inverse problem by first
 finding a canonical mapping from 
 the spectral data, which are like  action-angle variables,
 to  the scattering potential, i.e. the  original field and using it  
to construct  the exact solution for the
 given nonlinear equation. Soliton 
 is a special solution, which corresponds
 to the reflection-less 
 potential.
 
The QISM succeeds to generalise the notion of the 
 Lax operator also to the quantum case. Since the canonical variables 
 $u,p$ or $\psi,  \psi^\da$ etc. now  become  operators acting on the Hilbert 
space,
 the quantum $L(\la)$-operators 
 are unusual matrices with non-commuting matrix elements. This intriguing
feature leads to nontrivial underlying algebraic structures in QIS.
We present below a  few concrete examples
of the Lax operators associated with well known models to give  an idea about
the structure of this immensely important object 
in the  integrable theory. The field models are represented by field
equations and the corresponding continuum Lax operator ${\cal L}$, while the
lattice models are given through the Hamiltonian $H$ and the discrete Lax
operator $L$. 
\medskip

\noindent I. {\it Trigonometric Class}:\smallskip

\ni
1. Sine-Gordon (SG) model   

\begin{equation}
u(x,t)_{tt}- u(x,t)_{xx} = \frac {m^2}{\eta} \sin(\eta u(x,t)), \quad
  {\cal L}_{SG}  = \left( \begin{array}{c} ip , \qquad
  m  \sin (\la-\eta u) \\
   m  \sin (\la+\eta u),  \qquad -ip
    \end{array} \right), ~~p={\dot u}
\ll{sg}\end{equation}
                     \noindent
2. Liouville model (LM)    
\be
u(x,t)_{tt}- u(x,t)_{xx} = \frac {1}{2} e^{2\eta u(x,t)}, \qquad
  {\cal L}_{LM}  = i\left( \begin{array}{c} p , \qquad
   \xi e^{\eta u} \\
  \frac {1}{\xi}e^{\eta u},  \qquad -p
    \end{array} \right).
\ll{lm}\end{equation}

\noindent
3. Anisotropic $XXZ$ spin chain     
\bea
{ H} &=& \sum_n^N(\si_n^1 \si_{n+1}^1+\si_n^2 \si_{n+1}^2 +\cos \eta
\si_n^3 \si_{n+1}^3), \nonumber \\
{L_n }(\xi) &=&
 i \left( \begin{array}{c}
  \sin(\la + {\eta \ov 2}  \si_{n}^3) ,
 \qquad  \sin \eta \si_n^- \\
 \sin \eta \si_n^+    ,\qquad
  \sin(\la -  {\eta \ov 2}  \si_{n}^3) ,
    \end{array} \right)\ll{XXZ}\eea

\noindent II. {\it Rational Class}:
\sk

1. Isotropic $XXX$ spin chain     
\be
{ H} = \sum_n^N \sum_a^3(\si_n^a \si_{n+1}^a), \ \ \
{L_n }(\xi) =
  i\left( \begin{array}{c}
  \la +  {1 \ov 2} \si_{n}^3 
 \qquad   \si_n^- \\
 \si_n^+    \qquad
  \la - \si_{n}^3) 
    \end{array} \right)\ll{XXX}\ee

\ni
  2. Nonlinear Schr\"odinger equation (NLS)  

\begin{equation}
i\psi(x,t)_{t}+ \psi(x,t)_{xx} + \eta (\psi^\da(x.t)\psi(x,t))\psi(x,t)=0,
      \quad
{\cal L}_{NLS}(\la)  = \left( \begin{array}{c} \la ,
\quad \eta^\ha \psi \\
\eta^\ha  \psi^\da, \quad -\la
    \end{array} \right).
\ll{nls}\end{equation}

\ni
3. Toda chain   (TC)   
\be
H=\sum_i\left(\ha p^2_i
+ e^{(u_i-u_{i+1})}
\right),\qquad
L_n(\la) = \left( \begin{array}{c}
  p_n
-\la \qquad    e^{u_n}
 \\- e^{-u_n
}
 \qquad \quad 0
          \end{array}   \right).
\ll{tc}\end {equation}
Notice that we have put the models under  different classes
and apparently diverse looking models into the  
same  class.
 The meaning of this will be understood as we go further, though
  it might have  already sent a  signal  
 of the  fascinating unifying feature in the integrable systems.
Observing carefully we also see that 
 the off-diagonal elements (as $\psi, \psi^\da$ in (\re{nls}) and 
$\si^-,\si^+$ in (\re{XXZ})) involve {\em creation} and 
{\em annihilation}  operators while the diagonal terms are the number like 
operators. It is crucial to note 
 that   under matrix multiplication  this property
is preserved, which as we will see  below, has important consequences in their 
algebraic Bethe ansatz solution.

\section{Yang-Baxter equation, $R$-matrix and 
notion of  quantum integrability }
\setcounter{equation}{0}

As we see, the  Lax operators are  local functions of   $x$, or if we
discretise the space  of every lattice site $i$. However,
since the integrability is defined through the conserved quantities,
 which are
 global objects, in proving integrability we evidently need
  some global entries constructed from the local
  Lax operators.  
Such an object can be formed by matrix 
multiplication of the  Lax operators at all lattice sites as
\be
~~T(\la)=\prod_{i=1}^N L_i(\la)~=
 \left( \begin{array}{c}
  A(\la)
 \qquad    B(\la)
 \\ C(\la)
 \qquad D(\la)
          \end{array}   \right).
\ll{monod}\end {equation}
As we have indicated above, the off-diagonal
 global operators $B(\la), C(\la)$ are like creation/annihilation 
operators and related to the   angle-variables, while $ 
A(\la), D(\la)  $ correspond to  action variables.
 For ensuring integrability one must show   that 
$~\tau(\la)=tr T(\la)~=A(\la)+ D(\la)$
generates the conserved operators : $\ln \tau (\la)=\sum_j C_j
\la^j$ with  $~ [ H, C_m]=0$ and  
mutual commutation: $
 [ C_n, C_m]=0 .$   This is in fact achieved by a key condition
on the quantum Lax operators (for  
ultralocal models) given by 
\be~~ R_{12}(\lambda - \mu)~  L_{1i} (\lambda)~  L_{2i}(\mu )
~ = ~  L_{2i}(\mu )~ L_{1i }(\lambda)~~ R_{12}(\lambda - \mu).
\ll{qybel}\ee~
with the notations $~L_{1i}=L_i \ot I,~ L_{2i} =I \ot L_i. ~~$
This 
matrix relation is known as the Quantum Yang--Baxter equation (QYBE), where 
apart from $L$-operator   $R_{12}(\la,\mu)$-matrix  with c-number functions of
spectral parameters
and acting nontrivially only in the first two spaces $V\ot V \ot I$
 appears, which  in turn  satisfies the YBE
\be~~ R_{12}(\lambda - \mu)~  R_{13} (\lambda- \gamma)~  R_{23}(\mu -\gamma)
~ = ~  R_{23}(\mu -\gamma)~ R_{13}(\lambda -\gamma)~~ R_{12}(\lambda - \mu).
\ll{ybe}\ee
Due to
an additional ultralocal property of the $L$-operators:
 $ [L_{1i} (\lambda),  L_{2j} (\mu)]=0, i \not =j$ 
of a large class of  models, one can treat these 'specially separated'
operators almost as  classical objects. Therefore  multiplying them for all
sites and repeatedly using (\re{qybel}) one may  arrive at the  
 same QYBE
\be~~~ R_{12}(\lambda - \mu)~ { T_1} (\lambda)~ { T_2}(\mu )
~ = ~  { T_2}(\mu )~ { T_1 }(\lambda)~~ R_{12}(\lambda - \mu),\ll{qybeg}\ee
 but for the   global object $T(\la)$ defined in (\re{monod}).
This invariance relation for the tensorial product 
 reflects a   deep algebraic property related to the Hopf algebra,
to which we will return later. 
Taking now the trace of  relation (\ref{qybeg}), ( since under the trace
$R$-matrices can rotate cyclically and thus cancel out)  one gets  
$ [\tau(\lambda),\tau(\mu)]=0 ,$ proving 
the commutativity of $C_n$ for different $n$'s. 
Thus  starting from the local QYBE
(\re{qybel}) and following  several logical steps
 we finally establish 
 the complete  integrability of the quantum system. Therefore the validity 
of 
(\re{qybel}) may be considered to be  sufficient  for 
the quantum integrability of the ultralocal systems
 and  the local QYBE as the basic equation in the QIS. 
 We will show below that even for nonultralocal 
models  one can formulate the local QYBE in some generalised form 
 and prove the integrability, though 
the procedure is much more involved.
The  $4 \times 4\  $ $R(\lambda)$-matrix 
 solution of YBE \re{ybe} is rather easy to
find, which in    the simplest form may be    given as
\begin {equation}
R(\lambda) = \left( \begin{array}{c}
f(\lambda) \ \qquad \ \qquad \ \qquad \\
    \quad \ 1 \  \ \ \ f_1 (\la) \ \quad  \\
     \qquad \ f_1 (\la)\ \ \ \ \ \
     1 \ \quad \\
        \qquad \ \qquad \ \qquad \ f(\lambda)
          \end{array}   \right).
\ll{R-mat}\end {equation}
There  are usually only two  types of solutions 
(we shall not speak here of more general elliptic
 solutions), namely  {\it trigonometric} 
with
\be
 f(\la)={ \sin(\la+\eta) \ov   \sin \la },
~~~f_1 (\la) ={ \sin \eta \ov   \sin \la}  
\ll{trm}\ee
and the {\it rational}  with  \be
f (\la)= { \la+\eta \ov  \la},~~ 
 f_1 (\la) = {\eta \ov \la}  .
\ll{rrm}\ee
In the  examples of  integrable systems   the  trigonometric and
rational
classes mentioned are associated with the respective $R$-matrices 
presented here.

\section{ Universality in underlying algebraic structure}
\setcounter{equation}{0}
In sect. 2  we have given the Lax operators for a number of models 
 and grouped some like SG, LM and XXZ  in
the same trigonometric class, while equally varied models like NLS, TC,
$XXX$  etc. 
are in the same rational class. Here we address to the
 intriguing question that
 {\it
why} a wide range of   diverse models share the same $R(\la)$ and
 construct also their Lax operators  exploring the
universality feature underlying such systems.

To generate  models of the trigonometric class we start 
with the $R_{trig}$-matrix  solution (\re{trm}) 
  and look for the  the 
  Lax operator of the corresponding 
{\it ancestor} model in a generalised form 
\be L_t(\la)= i \left( \b{array}{c}\sin (\la+ \eta s^3) (c^+_1+c^-_1)-i
\cos (\la+ \eta s^3) (c^+_1-c^-_1),
\qquad  (2\sin \eta) S^- \ \\
(2\sin \eta ) S^+ , \qquad \sin  (\la -\eta s^3)(c^+_2+c^-_2)-i
\cos (\la+ \eta s^3) (c^+_2-c^-_2) \e{array}
\right), \ll{nlslq2}\ee
  The abstract operators $
 s^3,S^{\pm} $ describe  an algebra  given by the following relations
\be
 [s^3,S^{\pm}] = \pm S^{\pm} , [ S^ {+}, S^{-} ] = - {1 \over 2} 
 \left ( M^+[2 s^3]_q +i {M^- \over \sin \eta } \cos
( 2 \eta s^3  ) \right).
\ll{nlslq2a}\ee
where $  [x]_q= \frac {q^x-q^{-x}}{q -q^{-1}}
=  \frac {\sin( \eta x)}{\sin \eta} , \ \ q = e^{i \eta} 
 $ and $ M^\pm$ are related to the central elements of the algebra :
 $c^\pm_i, i=1,2$ as $ M^\pm= c^+_1c^-_2 \pm  c^-_1c^+_2 .$
 Note that (\re{nlslq2a}) is 
a new type of algebra underlying such 
 QIS. This is in fact
 a quadratic algebra and  unlike the known Lie algebra 
or its deformations there are arbitrary  multiplicative  
central elements  $M^\pm$  in
the RHS of (\re{nlslq2a}), which can take trivial values.
For different values of 
 these elements one gets different kinds of the 
algebra along with different types of the Lax operators.
 Therefore starting from this universal structure of the 
ancestor model (\re{nlslq2})  and considering various representation 
of (\re{nlslq2a}) with concrete choices for $c^\pm_i,$ 
one can generate systematically   the Lax
operators of diverse models belonging to this class. 

Remarkably, for the  choice  $ M^-=0, M^+=-2 $ one gets      
  the well known quantum algebra (QA)
 $U_q(su(2))$ with the defining relations 
\be
 [s^3,S^{\pm}] = \pm S^{\pm} ,\ \
  [ S^+, S^-] =   [2 s^3]_q 
\ll{sl2qa}\ee
Let us consider first the models related directly to this QA 
 with the Lax operators obtained from (\re {nlslq2})
for the choice $ c^+_1=c^+_2=-c^-_1=-c^-_2=1$.
 The  fundamental representation 
\be
S^\pm=\ha \si ^\pm, s^3=\ha \si ^3
\ll{qasc}\ee
constructs clearly  the Lax operator of the spin-$\ha$ XXZ-chain (\re{XXZ}).
Another nontrivial example obtained through the bosonic representation 
( with $[u_n,p_m]= {i \hbar \over \De}$) of 
QA:
\be
 s^3_n=u_n, \ \  S_n^-= g(u_n) e^{i \Delta p_n}, S^+_n=(S^-_n)^\dag
,\ll{qalsg}\ee
where $\ g(u_n) = [1+ \ha m^2 \Delta ^2 \cos 2 \eta (u_n+\ha)]^\ha \ $
yields an exact lattice version of the  sine-Gordon 
model. The Lax operator of the model can be 
derived readily from  (\re {nlslq2}) using (\re{qalsg}). 
 At $\De \rw 0$ one gets the SG field model with the 
Lax operator obtained as $\si^1 L_n=I+\De {\cal L}_{SG} (x) +O(\De) $ reproducing
 (\re{sg}).

We  now go beyond QA and consider  more general structure 
 (\re {nlslq2a}). Note that the choice
    $ c^+_1=c^-_2=\De, \ c^-_1=c^+_2=0 $ giving
$ M^+=M^-=\De^2$ corresponds to an intriguing  algebra with $
[ S^+, S^-]= i\De{q^{-2 s^3} \over \sin \eta }.$
  A bosonic realisation of this algebra
as 
\be
 s^3_n=u_n, \   S^+_n= f(u_n) e^{i \Delta p_n}, S^+_n=(S^-_n)^\dag
,\ll{qallm}\ee
with $f(u)= \left(1+\De^2 e^{i \eta (2u+i \hbar )}\right)^\ha$
constructs again from the same   (\re {nlslq2}) the Lax operator 
of an exact lattice Liouville model and at $\De \rw 0$ the 
Liouville field model (\re{lm}).
We  have witnessed here    an important consequence of the 
 the underlying algebraic structure, from which  through simple
realisations we could generate   exact lattice versions of  SG and LM, which
were originally found in an intuitive and much involved way  
\c{lsg}\c{llm}.
 
We have shown  thus that  
 the Lax operators of the   models, at least for 
 the trigonometric class
 presented in sect. 2, can be constructed  
systematically and  the set of diverse models  put under the same
class are in fact descendants of the same ancestor model 
 (\re {nlslq2}). We continue this procedure to 
generate other members of this class 
like the quantum Derivative NLS (DNLS), Ablowitz-Ladik model,
relativistic Toda chain etc.
as different 
realisations of the unifying algebra (\re {nlslq2a}).
For example, at 
$ c^+_1=c^+_2=1, \ c^-_1=i {\De \over 4} q , \ c^-_2= i {\De \over 4 q} $
corresponding to $ M^+= {\De \over 2} \sin \eta
, \ M^-= i {\De \over 2} \cos \eta, $ the quadratic algebra (\re {nlslq2a}) 
after the substitution
$$ S^+= \kappa A, \ S^-= \kappa A^\dag, \ s^3= N, \ \ \kappa= -i (\cot \eta)^\ha
$$ leads directly to the well known $q$-oscillator algebra
\be
 [A, A^\dag]= {\De \over cos \eta} \cos (\eta (2N+1)), [A,N]=A,
A^\dag,N]=-A^\dag
,\ll{qoscl}\ee
 and therefore to a novel integrable $q$-oscillator model with explicit 
 $L$-operator and the $R_{trig}$ -matrix (\re{trm}).
 A realisation of the $q$-oscillators in
 bosonic operators: 
$ [\psi_n , \psi_m^\dagger ] = {\hbar \over \Delta }\de_{nm}$
in the form
  \be  A_n =
     (\frac{\Delta}{\hbar})^\ha~\psi_n \sqrt\frac{[N_n]_q}{N_n}, \ N_n=
   \frac{\Delta}{\hbar}  \psi^\da_n\psi_n \ll{qo-b}\ee
derives an interesting  exact lattice version of the
 quantum DNLS
model, which at the  $~~\lim_{\De \rw 0 }~$ yields the Lax operator
\begin{equation}
  {\cal L}^{dnls}(\xi)  = i
  \left( 
  -\frac{1}{4}\xi^2 \si^3  +\psi^\da(x)\psi(x) {\bf \kap }
 + {\xi}(\psi^\da(x) \si^+ + \psi(x) \si^-) 
     \right),~~ {\bf \kap}= diag (\kappa_-, -\kappa_+)
\ll{dnls1}\end{equation}
of  the DNLS field model,
which is  exactly solvable through the Bethe ansatz \c{qdnls}.
A fusion of such DNLS model in turn can generate the well known massive
Thirring model \c{construct}.

A different set of choice for the elements $c_i^\pm$ likewise results 
from (\re {nlslq2a})
other algebras and related  integrable models. The simplest choice
$c_i^-=0, c_i^+=1 $  leads to a light-cone SG model, while 
$c_2^\pm=0, c_1^-=-c_1^+=1 $ giving $M^\pm=0$ with  realisation
\begin{equation}
s^3= p, \ S^-={\eta \over \sin \eta} e^u, \ 
S^-={\eta \over \sin \eta} e^u
\ll{rtoda}\end{equation}
 gives from the ancestor model
(\re {nlslq2a}) the Lax operator of a quantum relativistic Toda chain 
\c{rtoda}.
It should also be noted that under a { twisting} transformation: 
\be
R^\th_{trig}(\la)= F(\th) R_{trig}(\la) F(\th), \ \ \mbox{with} \ \ 
F(\th)=
e^{i\th(\si^3 \ot 1- 1 \ot \si^3)}\ll{twist}
\ee
 one gets a twisted 
trigonometric $R$-matrix and a corresponding $\th$-parameter 
extension   of the algebra (\re {nlslq2a}). Proceeding as above  and 
choosing properly $\th$ one
gets yet another set of integrable models, a prominent example of which is
the Ablowitz-Ladik model \c{ALM}\c{construct}.  
This demonstrates clearly the unifying feature of the algebraic 
structure in generating a wide range of descendant models from the  
Lax operator (\re {nlslq2}) all sharing the same trigonometric $R$-matrix
inherited  from the ancestor model.

Let us consider now the
 $q \rightarrow 1$ limit  when  $R_{trig} \rightarrow R_{rat} $ giving
  the rational $R_{rat}$-matrix 
 (\re{rrm}), which can  be expressed also as $R_{rat}(\la)=
\la+\hbar P$. The ancestor model  reduces to the corresponding
rational form
\begin {equation}
L(\lambda)_{rat} = \left( \begin{array}{c}
{K_1^0 + i  { \lambda  } c_1^0 }
\qquad   \ \  {K_{21}}
  \\     {K_{12}}
 \ \ \qquad  {K_2^0 + i  { \lambda } c_2^0 }
          \end{array}   \right).
\ll{l-anc0}\end {equation}
As a consequence of  the QYBE
(\re{qybel}) the $ {\bf K }$ operators  satisfy  the 
algebra given through the relations
\[  [ K_{12} , K_{21} ]
=  (c_2^0 K_1^0  - c_1^0 K_2^0 ) , ~~ [ K_1^0, K_2^0 ] = 0
              \]  
\be       [K_i^0, K_{12}] = 
 \ep_i K_{12} c_i^0 , ~ [K_i^0, K_{21}]  = - \ep_i K_{21} c_i^0,
  \ll{k-alg} \ee
with $\ep_1=1, \ep_2=-1$ and  $ c_1^0, c_2^0 $ are the central elements.
Note that though it is much similar to a Lie algebra, it is in fact again
a quadratic algebra with arbitrary multiplicative  central elements. 
Since    $L(\la)$ operator given in the general form  (\re{l-anc0})
satisfies QYBE and associated with the rational 
$R_{rat}$-matrix,
 it may be taken as   the  ancestor model
for generating  
  integrable  models belonging to the rational class.

A nontrivial choice for the central elements $ c_1^0 =c_2^0  =  1$ and the
notation 
\be ,
~~ K_1^0  =  -K_2^0= s^3,  ~~ \  K_{12}=  \ s^+, \    K_{21}=  \ s^-
\ll{K-s}\ee
 recovers   the    $su(2)$ spin  algebra
\begin {equation}
 [s^3,s^{\pm}] = \pm s^{\pm} ,\quad
  [ s^+, s^-]= 2 s^3,
\ll{sl2a}\end {equation}
which is a Lie algebra.
The integrable models like  the $XXX$ spin chain and the NLS  model,
our acquaintances  from sect. 2, can be generated from the case (\re{K-s}).
In fact the simplest 
 spin-$\ha$ representation through Pauli matrices $s^a= \ha \sigma^a$ 
 reduces (\re{l-anc0}) to 
 the Lax operator of 
the $XXX$ chain  (\re{XXX}). On the other hand 
 bosonic representation of the spin operators given by the 
Holstein-Primakov transformation
\be 
s^3= s- \De \psi^\da \psi, s^-=    \De^\ha(2 s- \De\psi^\da
\psi)^\ha\psi^\da, \ \ s^+= (s^-)^\da
\ll{hpt}\end {equation}
leads 
 (\re{l-anc0}) to the quantum integrable 
 Lattice NLS model. The continuum limit takes it to the NLS field model
(\re{nls}).

 The general algebra (\re{k-alg}) permitting trivial values for $c_i^0$ 
however allows 
more   freedom for constructing models. Choosing $c^0_2=0, \ c^0_1=i$
for example  one can set
  $ ~ K_2^0  = 
=0 , $  reducing the algebra to
 \be [K_1^0 , K_{12}] =  i K_{12},~~[K_1^0 , K_{21}] = - i K_{21}
~ [K_{12} , K_{21} ] = 0 ~.\ll{altoda}\ee
A   bosonic representation   of the generators:  \be  K_1^0
=  p , K_{12} = e^{-u},~~K_{21} = e^{u}, \ll{k-b}\ee
  as is evident
  from 
(\re{l-anc0})  yields
the Lax operator of the  Toda chain (\re{tc}).
Another simple lattice NLS model can be obtained 
taking again  the values as $c^0_2=0, \ c^0_1=-\De \kappa^2$  
 with the representation
\be K^0_1=\kappa \Delta^2 \phi \psi+1,~~ K_2^0= 1, ~~ 
 K_{12} = i \De \sqrt{\kap} \psi,~~K_{21} = - i \De \sqrt{\kap} \phi
 \ll{kslnls-b}\ee
where the operators $~\psi, \phi~$ are canonical.
At the continuum limit the same standard NLS field model (\re{nls}) is recovered.

Finally we can consider the  twisted rational $R^{\th}_{rat}(\la)$-matrix
through  transformation  (\re{twist}) with the same twisting
operator
$F(\th).$  As a result one gets 
 a $\theta$-deformation 
  of the   algebra  (\re{k-alg}) in which  
   $c^1_0$'s  no longer remain central elements.
We can construct again in a similar way the integrable models belonging to
this class. Thus one can obtain 
the  lattice $\th$-deformed NLS, $\th$-deformed    Toda chain, 
Tamm-Dancoff $q$-bosonic model or Dzhyaloshinsky-Moriya 
spin chain, which is nothing other than the $\th$-deformed $XXX$ chain
\c{construct}\c{kunrag}

Thus the algebraic structure in the QIS clearly plays a crucial role 
in 
 unifying  diverse models of the same class as descendants from the same
ancestor model and at the same time realisations like (\re{qalsg}) gives a
criterion for defining {\it integrable nonlinearity} as different nonlinear
realisations of the underlying quantum algebras
 (\re {nlslq2}) or (\re {k-alg}). We  see below that this universality
feature is also present curiously in their Bethe ansatz solution, which
facilitates their exact eigenvalue solution  in an almost model-independent
way. 
 
\section{Universality in     Bethe ansatz  solutions}
\setcounter{equation}{0}

Coordinate formulation of the Bethe ansatz, though more effective for 
finding the energy spectrum :$  H\mid m>= 
  E_m\mid m> $ of concrete models,  dependents heavily on the structure 
of the Hamiltonian  and consequently  lacks the uniform approach
of its algebraic formulation.
We therefore focus briefly only  on the   algebraic Bethe ansatz 
method for    solving the general  eigenvalue problem 
\be
  \tau(\la)\mid m>= 
  \Lambda_m(\la)\mid m>
\ll{evp}\ee
 and highlight its universal feature.

Apart from the integrability condition, 
the QYBE (\ref{qybeg}) represents also   a set of 
 commutation relations between  {\it action} and {\it angle} variables
given in the matrix form. They 
 can be obtained by inserting   
matrix $T$ in the form  (\re{monod})  and
the     $R(\la)$-matrix solution as (\re{R-mat}) in the equation 
 (\re{qybeg}).
Such generalised commutation relations dictated by the QYBE are of the form
\bea
A(\la) B(\mu)&=&  f(\mu-\la)  B(\mu) A(\la)  - f_1(\mu-\la)  B(\la) A(\mu)
 , \ll{ab} \\
D(\la) B(\mu)&=&  f(\la-\mu) B(\mu) D(\la) -  f_1(\mu-\la)  B(\la) D(\mu), \ll{db} 
\eea
together with the trivial commutations for  $ [A(\la), A(\mu)]  = 
[B(\la), B(\mu)]  = [D(\la), D(\mu)]  = [A(\la), D(\mu)]  = 0$ etc.
It is now important to note that the $off$-diagonal element $B(\la)$ acts
like an creation operator (induced by the local creation operators of 
 $L(\la)$ as argued above). Therefore
the $m$-particle state $\mid m>$ may be considered to be created 
by $B(\la_i)$ acting $m$ times on the pseudovacuum $\mid 0>$:
 \be \mid m>= B(\la_1) B(\la_2)\ \cdots B(\la_m)\mid 0>
\ll{phim}\ee
If one can  solve
 the general 
problem (\re{evp}),
the eigenvalue problem for all  
\be
C_n= {1 \ov n!} {\partial
\over  \partial \la}\tau_N(\la)\mid_{\la=0}
\ll{cn}\ee 
can be obtained 
simultaneously by simply  expanding
$\Lambda(\la)$ as
\be
  C_1 \mid m>= 
  \Lambda_m'(0)\Lambda_m^{-1}(0)\mid m>
,~~C_2 \mid m>= 
  (\Lambda_m'(0)\Lambda_m^{-1}(0))'\mid m>
\ll{evpcn}\ee
etc. , where say the Hamiltonian  $H=C_1$.

Evidently for solving (\ref{evp}) through the Bethe ansatz we have to drag $
\tau(\la)=A(\la)+ D(\la)$ through the string of $B_{\la_i}$'s 
without spoiling their structures (and thereby preserving the eigenvector)
and hit finally the 
pseudovacuum giving $A(\la)\mid 0>=\al(\la)\mid 0>$ and $
D(\la)\mid 0>=
\bet(\la)\mid 0>$.

Notice that for this purpose (\re{ab},\re {db}) coming from the QYBE  
are indeed the right kind of commutation relations but for the second
terms in both the RHS, where the argument of $B$
  has changed: $\ti B \rw B$
spoiling the structure of the eigenvector. However, if we put  the
sum of all such {\it unwanted} terms= 0, we should be able to achieve our goal. 
In field models such unwanted terms are  however  absent, while in lattice
models
they may be removed
by the  Bethe equations induced by
the periodic boundary condition giving
\be 
\l({ \al(\la_k) \ov \bet(\la_k)}\r)^N=
 \pr_{l \not =k}{ f(\la_k-\la_l) \ov
 f(\la_l-\la_k)}
= ~ \pr_{l \not =k}{ -a(\la_k-\la_l) \ov
 a(\la_l-\la_k)}, ~~ k=1,2, \ldots ,m.
\ll{be}\ee
 This in turn serves as the 
  determining equations for   parameters $\la_j$.
 Ignoring  therefore  the second terms for the time being and making  
use  of the first terms only, as argued  above 
 we finally solve the eigenvalue problem to yield
\be \Lambda_m (\la) =
 \pr_{j=1}^m f(\la_j-\la)
\al(\la)+
\pr_{j=1}^m f(\la-\la_j)
\bet(\la).
\ll{lambda}\ee

The structure of the eigenvalue $\Lambda_m (\la)$  reveals the curious fact
about the Bethe ansatz result,
 that apart from the $\alpha(\al),\bet(\al)$ factors 
the eigenvalue (\re{lambda}) and the Bethe equation (\re{be})
 depend mainly  on the nature of the function $f(\la-\la_j)$, which
 are universal within a class of models and determined only by the choice of 
$R$-matrix as 
  (\re{trm}) or (\re{rrm}).
   The coefficients  $\al(\la), \bet (\la)$ 
determined by the concrete form of the Lax operators 
and the definition of the  pseudovacuum are the only 
model-dependent  part. 
Therefore  models like the DNLS, SG, Liouville
and the $XXZ$ chain belonging to the trigonometric
 class share  similar  type of
eigenvalue relations with individual differences only in 
the   form of  $\al(\la)$ and $ \bet (\la)$ coefficients.
Thus  
this  deep rooted
 universality  in integrable systems  helps to solve the eigenvalue
problem for the whole class of models and for the full  hierarchy of
their 
conserved currents in a systematic way. 
For example, for the $XXZ$ model
using the Lax operator structure (\re{XXZ}) and
 $\mid 0>$ defined as the all spins up
state, one easily finds 
$\al(\la)=\sin ^N (\la+ \eta), \ \bet(\la)=\sin ^N \la $. Using subsequently
the $R_{trig}$-matrix information (\re{trm}) one derives 
 from (\re{be}) (with  a shift $\la \rw \la+{\eta \ov 2} $)
 the Bethe equation 
\be 
\left(  {\sin (\la_k+{\eta \ov 2}) \ov \sin (\la_k -{\eta \ov 2})} \r)^N=
 \pr_{j \neq k}^m {\sin (\la_k-\la_j+\eta) \ov \sin (\la_k-\la_j - \eta)}.
\ll{bexxz}\ee
for $j=1,2, \ldots , m.
$   Similarly (\re{lambda}) gives  the eigenvalue 
\be \Lambda_{XXZ} (\la) =\sin ^N (\la+ \eta)
 \pr_{j=1}^m {\sin (\la_j-\la+{\eta \ov 2}) \ov \sin (\la_j-\la -{\eta \ov 2})}
 +\sin ^N \la
\pr_{j=1}^m {\sin (\la -\la_j+3{\eta \ov 2}) \ov \sin (\la-\la_j+{\eta \ov 2})}
\ll{lamxxz}\ee
 yielding for 
 $H_{xxz}=C_1,$   the energy
spectrum
\be
  E_{xxz}^{(m)}= \Lambda (\la)'\Lambda^{-1} (\la)\mid_{\la=0}=
\sin \eta\sum_{j=1}^m {1 \ov \sin (\la_j -{\eta \ov 2}) 
\sin (\la_j + {\eta \ov 2})}+ N\cot \eta.  
\ll{exxz}\ee

At the  limit $\eta \rw 0, \sin \la \rw \la$, when the $R$-matrix along
with its associated models reduce to the rational class, one can derive 
the Bethe ansatz result of the isotropic $XXX$ chain directly 
by taking the limits of the above expressions for $XXZ$ model. The Bethe
ansatz  
 result for the  NLS
model of the same rational class also have very similar structure.

It should be remarked that unlike the coordinate formulation, the algebraic
Bethe ansatz does not require explicit form of the Hamiltonian. The
information about the Lax operator and the $R$-matrix is more important for
this method. However using the definitions (\re{cn}) and (\re{monod})
we can construct Hamiltonian of the model from the Lax operator. For
example,
 the $XXZ$ chain with  Lax operator (\re{XXZ}) ( with a shift $\la \rw
\la + {\eta \ov 2}$) satisfies the condition
 $L_{ai}(0)=P_{ai}=I+ {\bf \si \cdot \si} $, where $P_{ai}$
is the permutation operator with the property $P^2=1$ and $P_{ai}
L'_{ai+1}(0)=L'_{ii+1}(0)P_{ai}$. 
Using this property we derive from (\re{monod})
\be
\tau (0)=tr_a\l(P_{aj} P_{aj+1}\ldots  P_{aj-1}\r) 
\ =(P_{jj+1}\ldots  P_{jj-1})tr_a(P_{aj}) 
\ll{tau0}\ee
for any $j$, applying freedom of cyclic rotation of matrices under the 
trace.
Taking derivative with respect to $\la$ in (\re {monod}) we similarly get
\be
\tau' (0)=tr_a \sum_{j=1}^N 
\l(P_{aj} L'_{aj+1}(0)\ldots  P_{aj-1}\r) \ \ = \
\ \sum_{j=1}^N(L'_{jj+1}(0)\ldots  P_{jj-1})tr_a(P_{aj}) ,
\ll{tau1}\ee
where we have assumed the periodic boundary condition: $
 L_{aN+j}= L_{aj} $. Defining  now from (\re{cn}) 
 $H=c~ C_1= c{d \ov d \la} \ln \tau(\la)_{\mid \la=0}
 =c~ \tau' (0)\tau^{-1}(0),~c  =const.$
 we get the Hamiltonian  from   (\re {tau0}) and 
(\re {tau1}) as 
\be
H=c \sum_{j=1}^N(L'_{jj+1}(0) P_{jj+1})
\ll{htau01}\ee
with only $2$-neighbor interactions, 
due to cancelation of all other nonlocal factors and the property  $P^{-1}=P$.
Calculating $L'$ from the Lax operator (\re{XXZ}) and inserting in
(\re{htau01})
we finally get  the explicit form of $H$ as in (\re{XXZ}) for the $XXZ$
chain.  
\section {Algebraic structures in  ultralocal and
nonultralocal integrable systems}
\setcounter{equation}{0}
We have seen that the algebraic property in integrable systems
 is dictated  by the QYBE. However since the QYBE  
for the  nonultralocal systems is different from ({\re{qybel}), the related
  algebraic structure  can not be described by the above procedure.
As a result  progress in this field  
is not much  satisfactory compared to the above described 
ultralocal theory, though many important models like 
Nonabelian Toda chain, quantum KdV and modified KdV model, nonlinear
$\si$-model etc. belong to this class.
 Nevertheless, 
it is possible now  to describe  such models to considerable extent 
by some extensions 
of the QYBE \c{maillet} \c{kunhla}. 

However for better understanding of the algebraic structure of the 
nonultralocal models let us take a closer look at that of the
ultralocal Lax operators themselves.
The underlying quadratic quantum algebra (\re{nlslq2a}),
 as mentioned before, exhibits Hopf algebra
property. The most prominent characteristic of it is the coproduct structure
given by 
\be
\De (s^3)= s^3 \otimes I+I\otimes s^3, \ \ \
\De (S^+) =  S^+ \otimes c^+_2 q^{-s^3}+ c^+_1 q^{s^3}\otimes S^+, 
\\
\De (S^-) =  S^- \otimes c^-_1 q^{-s^3}+ c^-_2 q^{s^3}\otimes S^-
\ll{copr}\ee
Note that the $c^\pm_i=1$ case recovers the well known result for the 
standard quantum algebra.
This means that  if $S^\pm_1=S^\pm \otimes I$  and 
 $S^\pm_2= I \otimes S^\pm$ satisfy the quantum algebra separately, 
then their tensor
product   $\De (S^\pm)$ given by (\re{copr}) should also satisfy
 the same algebra.
This Hopf algebraic property  induces the crucial transition  from the
local QYBE (\re{qybel}) to its tensor product given by the global equation 
(\re{qybeg}) as discussed above. 
Note that thanks to the property of the generators of the algebra: 
 $ [ S^a_{i},S^b_{j}]=0, \ i \not = j$ the ultralocality of the
Lax operators  
 is achieved.
However for  nonultralocal models  with 
  the underlying braided algebra having a different multiplication rule
though a similar Hopf algebra property \cite {majid},
we get  $[L_{1i}, L_{2j}] \not = 0$
 requiring generalisation of 
 the QYBE.

Such  generalised  QYBE for nonultralocal  systems
with the inclusion of  braiding matrices 
  $Z$ (nearest neighbour braiding)   and $\tilde Z$ (nonnearest neighbour
   braiding ) may be given by
\begin{equation}
{R}_{12}(u-v)Z_{21}^{-1}(u,v)L_{1j}(u)\tilde Z_{21}(u,v)L_{2j}(v)
= Z_{12}^{-1}(v.u)L_{2j}(v) \tilde Z_{12}(v,u)L_{1j}(u){R}_{12}(u-v).
\ll{bqybel}\end{equation}
In addition, this must be  complemented 
 by the  braiding relations  
\begin{equation}
 L_{2 j+1}(v)Z_{21}^{-1}(u,v)L_{1 j}(u)
=\tilde Z_{21}^{-1}(u,v)L_{1 j}(u)\tilde Z_{21}(u,v)
 L_{2 j+1}(v)\tilde Z_{21}^{-1}(u,v)
\ll{zlzl1u}\end{equation}
  at   nearest neighbour
points and
\begin{equation}
 L_{2 k}(v)\tilde Z_{21}^{-1}(u,v)L_{1 j}(u)
=\tilde Z_{21}^{-1}(u.v)L_{1 j}(u)\tilde Z_{21}(u,v)
 L_{2 k}(v)\tilde Z_{21}^{-1}(u,v)
\ll{zlzl2u}
\end{equation}
with $k>j+1$ 
  answering for the nonnearest neighbours.
Note that along with  the usual quantum $ R_{12}(u-v)$-matrix like
(\re{R-mat})  
additional  $ \  \tilde Z_{12} , \ Z_{12}$ matrices 
 appear, which can be (in-)dependent of the spectral parameters
and 
satisfy  a system of Yang-Baxter type relations
 \cite{kunhla}.

\section { Generation of nonultralocal
models from the  braided  QYBE}
\setcounter{equation}{0}

Unlike  ultralocal models
due to the appearance of $Z$ matrices in the braided QYBE 
relations   one faces  initial 
difficulty in trace
factorisation.
 Nevertheless, in nonultralocal  cases
 one can mostly  bypass this problem  by
introducing a $K(u)$ matrix and    defining 
$t(u)=tr (K(u)T(u))$ as commuting matrices  \cite{skly-r,kunhla} for
establishing the quantum integrability for nonultralocal models.
Though a well-framed theory for such systems is yet to be achieved,
one can
 derive 
  the basic equations for a series of nonultralocal models 
in a rather  systematic way from the  general relations
  (\ref{bqybel}-\ref{zlzl2u})
by 
 paricular explicit choices of  $Z,\tilde Z$ and $R$-matrices
\cite{kunhla,kunalush}. The models which can be covered through 
this scheme are 
\\ 
I. {\it Nonultralocal models with rational $R$-matrix}
\\
1. { Nonabelian Toda chain }\cite{natoda}

$\tilde Z=1, Z=I+i\hbar (e_{22} \otimes e_{12}) \otimes \pi.$
\\
 2. { Nonultralocal quantum mapping }\cite{Nijhof}

 $\tilde Z=1$ and $Z_{12}(u_2)
 = {\bf 1}+ \frac { h }{u_2}\sum_\alpha^{N-1}e_{N \alpha
}\otimes e_{\alpha N}~~~.$
\\
3. {\it Supersymmetric models } 

 $Z=\tilde Z=\sum \eta_{\al \bet} g_{\al \bet}$, where $ \eta_{\al \bet}= 
e_{\alpha \alpha}\otimes e_{\beta \beta}$ and $ g=
 (-1)^{\hat \alpha \hat
 \beta}$ with  supersymmetric grading  $\hat \alpha.$
\\
4. { Anyonic type SUSY model}

$Z=\tilde Z=  
 \sum \eta_{\al \bet} \tilde g_{\al \bet}$, with  
   $\tilde g_{\al \bet}=e^{i \theta 
\hat \alpha \hat
 \beta}$.
\\ 5. { Kundu-Eckhaus equation} \cite {kun84}

Classically integrable  NLS equation with 5th power nonlinearity  
\be i\psi_1+\psi_{xx}+ \kappa (\psi^\dagger \psi) \psi
+ \theta^2   (\psi^\dagger \psi)^2 \psi +2i \theta
 (\psi^\dagger \psi)_x \psi =0, \ee
as a  quantum model involves   anyonic type fields:
 $ ~\psi_n \psi_m= e^{i \theta} \psi_m\psi_n,~~ n>m;  
~~[ \psi_m, \psi_n^\dagger]=1.$ The choice
 $\tilde Z=1,
Z= diag (e^{i\theta},1,1,e^{i\theta})$ 
  constructs the braided QYBE,
 The trace factorisation problem
 has not been solved.
\\
\\
II. {\it Nonultralocal models with trigonometric $R$-matrix}
\\ 1. { Current algebra in WZWN model} \cite{wzwn}
 
 $\tilde Z=1$ and $R_{12}=Z_{12}=R_{q12}^-,$ where $R_{q}^\pm $ is the $\la \rw \pm
\infty $ limit  of the trigonometric $R(\la)$-matrix
\\
 2. { Coulomb gas picture of  CFT} \cite{babelon}
 
  $\tilde Z=1$, 
 $Z_{12}=
q^{-\sum_i H_i
\otimes H_i}
$ and $R_{12}=R_{q12}^+
$.
 \\ 3. {Integrable model on moduli space }\cite{alex}

  $\tilde Z=Z_{12}=R^+_q$ and $\la$-dependent $R(\la)$-matrix
 \\ 
 4. { Quantum mKdV model  }\cite{kmpl95}
 
 $\tilde Z=1, ~
Z_{12}=  Z_{21}= q^{-\frac {1}{2} \sigma^3\otimes \sigma^3
},$
and  the $\la$-dependent trigonometric $R(\la)$-matrix.

Other   models of nonultralocal class 
are the well known { Calogero-Sutherland} (CS) and 
{Haldane-Shastry} (HS) models with interesting long-range interactions.
Though spin CS model exhibits many fascinating features
 \c{hal-pas} and its integrability formulation  
through 
braided QYBE for both HS and CS models has not yet been achieved.

\section {Connections with Stat Mech and CFT }
As it was   mentioned above the unifying feature of    
the QIS  is  present  in its   
  relationships 
 with various other branches of physics and mathematics,
where the knowledge and techniques of the QIS  often proved to be  
extremely helpful.
Among these various connections those with the
  Stat Mech problems and the CFT models seem to be the most 
 interesting.

The $(1+1)$ dimensional quantum systems are  intimately linked  with $2$
dimensional classical statistical models. Moreover  the notions
 of integrability are 
equivalent in both these cases and are  through the Yang-Baxter relations.
 In   integrable statistical systems
 both QYBE (\re {qybel}) and  
 the  YBE (\re {ybe}) become the same and leads similarly to the commuting
transfer matrices $\tau(\la).$   
A classical  statistical  vertex model may be given by  
  $N\times M$
lattice points connected
by the bonds assigned with +ve (-ve) signs or equivalently, with right, up
(left, down) arrows  in a random way (see Fig. 1).

\par
\epsfxsize=6in
\epsfysize=3in
\epsffile{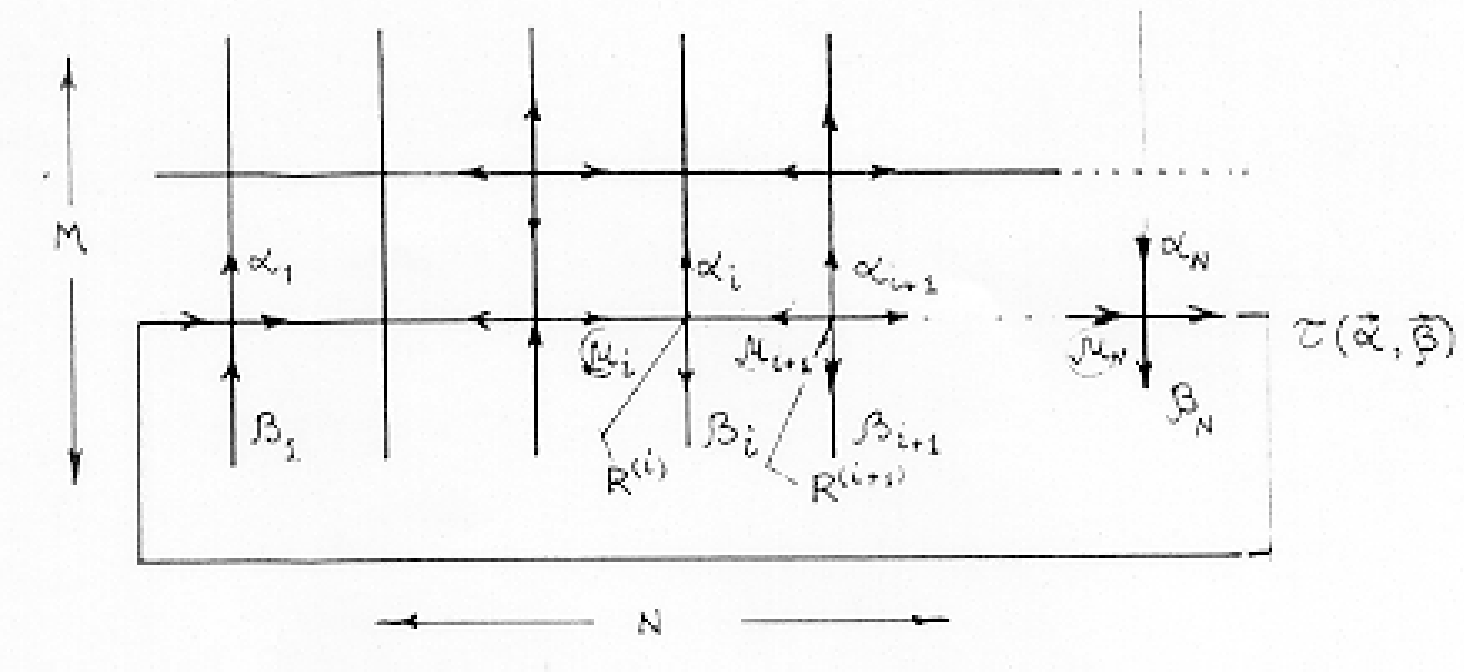}
\par

\vspace*{ 1cm}

  {\bf
Figure 1 } \ {\it Classical $2$-dimensional vertex model
is related with  $(1+1)$-quantum system. The row-to row transfer matrix
$\tau$ depends only on vertical space indices $\al, \bet$ and 
made out of multiplying $R^{(i)}$-matrices with elements 
$R^{(i)kl}_{\  \ mn}$, which represent
 the Boltzmann weights at vertex point
$i$.}

 Setting the corresponding Boltzmann
weights for  possible
arrangements 
 as  matrix elements of a
 $4\times4$-matrix, we get the $R^{(i)}_{12}
$-matrix with crucial dependence  
 on  spectral  parameter $\la$.
The YBE (\re {ybe}) restricts the solution of the $R$-matrix 
 to  integrable models.
Imposing   extra symmetries  
 on the $R$-matrix like
 the  { charge} conserving: $R^{ij}_{kl} \neq 0, ~$, only  when 
$k+l=i+j$ and  reversing:$R^{ij}_{kl}=R^{-i.-j}_{-k,-l} $ 
symmetry  we get the 
  Boltzmann weights  of the $6$-vertex model constituting the
elements of the $R$-matrix:
$$
 R^{++}_{++}=
R^{--}_{--}=
 a(\lambda), \ \ 
R^{+-}_{+-} =
R^{-+}_{-+}=
 b(\lambda), $$
 $$
R^{+-}_{-+} =
R^{-+}_{+-}=
 c(\lambda),$$
where ${a \over b}=f
,  {c \over b}=f_1. $ 
 Using an overall normalisation  one 
immediately recognises this as  our familiar
  $R$-matrix  (\ref{R-mat}), which 
represents also the Lax operator of the anisotropic quantum
 spin-$\ha$ chain (\re{XXZ}).

The configuration 
 probability for a string of $N$-lattice 
 sites in a row may be given by  the transfer
matrix 
$ \tau_N(\vec {\alpha},\vec{\beta})
= tr (\prod_i^N R^{(i)})$ in a similar way as in QIS.
The free parameter $\la$ 
may be linked with the {\it temperature} $T= {1 \over
k_B \bet} $  of the
statistical model  as $ \ \la= {const. \ov M} \bet . $
The direction along $N$ in statistical system   may be 
considered as the {\it space} direction of the quantum model,
 while the $M$ direction is like  the {\it time} direction  in the quantum
system  (see Fig. 1), which again 
 may be replaced in statistical systems by 
the temperature. Therefore, 
the finite-temperature behaviour can be studied by taking suitably
 the limit of $M$ \c{wadft}.   
Recalling that the Hamiltonian 
of the quantum model may be given by $ H=C_1= -\tau_N(0)^{-1}{ \partial
\over  \partial \la}\tau_N(\la)\mid_{\la=0}$, we can establish the
important 
connection $ exp(-\bet H)= \lim_{M \rw \infty} \left(\tau_N^{DTD} \right)^M$,
where $\tau^{DTD}_N=\tau_N(0)^{-1}\tau_N( {\bet \ov M}) $ is the diagonal to
diagonal transfer matrix. On the other hand  the partition function 
 $Z_{N.M}(\la)=  tr\left(\tau_N^{DTD} (\la) \right)^M$. Interchanging $N,M$ amounts
to the rotation by $90^o$ and effected by changing the spectral parameter
$\la \rw \eta -\la $ due to the crossing symmetry of the model.
This gives therefore
  $Z_{M.N}(\la)=  tr\left(\tau^{DTD}_M( \eta -\la) \right)^N$ yielding
the free energy $f=-{1 \ov \bet} \lim_{M \rw \infty} \Lambda_M(\la)$,
where $\Lambda_M(\la)$ is the largest eigenvalue of $\tau^{DTD}_M( \eta -\la)
 $.Therefore   using the eigenvalue expressions (\re{lambda}) and expanding 
it in powers of $\bet$ 
one can find the finite-temperature corrections to the thermodynamical
quantities just as in the case of expansion in large $N$  for finding the 
finite-size corrections as shown below.

Similar  deep interrelation exists also 
between the QIST and the CFT models,
 revealed first perhaps by Zamolodchikov \cite{zamcft} by   showing that, 
if CFT is perturbed through relevant perturbation and the system goes away
from criticality, it might generate hierarchies of integrable systems. For 
example $c=\ha$ CFT perturbed by the field $\si=\phi_{(1,2)} $ as
$H=H_{\ha}+h \si \int \si(x) d^2x, $ represents in fact the Ising model at
$T=T_c$ with nonvanishing magnetic field $h$. Similarly  the 
WZWN model perturbed by the operator $  \phi_{(1,3})$ generates  integrable
restricted sine-Gordon (RSG) model.
Under such perturbations the trace of the tress tensor, unlike pure CFT
 becomes nonvanishing and generates  infinite series of
integrals of motion associated with the integrable systems.

Another practical application of this relationship 
 is  to extract  important information about
the underlying CFT  in the scaling limit  of the
integrable lattice models. Interestingly, from the finite size correction
of the Bethe ansatz solutions, one can determine  the CFT
characteristics like the central charge and the  conformal dimensions
 \c{vegacft}.
For example, 
one may analyse the finite size effect of the Bethe ansatz solutions
of the $6$-vertex model ( with a seam 
given by  $\kappa$). Considering the coupling parameter 
 $q=e^{i{ \pi \ov \nu+1}},$ one obtains from the Bethe solution 
 at the large
$N$ limit  the expression
\[ E_0=Nf_{\infty}-{1 \ov N}{\pi \ov 6} c + O({1 \ov N^2}) \]
 for the ground state energy
and
\[ E_m -E_0 ={2 \pi \ov N}( \De+ \ti \De) + O({1 \ov N^2}) \qquad
 P_m -P_0 ={2 \pi \ov N}( \De- \ti \De) + O({1 \ov N^2}) \]
 for the excited states. Here 
  $\De,  \ti
\De$  are conformal weights of  unitary minimal
models  and 
$c=1-{6 \kappa^2 \ov \nu(\nu+1)}, \ \nu=2,3, \ldots$
is the central charge of the  corresponding conformal field theory.  

\section {Concluding remarks}
The variety of models within the   quantum integrable systems
are linked together by an unifying algebraic structure induced by
the quantum Yang-Baxter equation. Such structures not only  
systematically generate different classes of integrable models 
but also gives  an unifying approach in their Bethe ansatz solutions.
The theory of nonultralocal models, though not so successful allows to
derive the models from few basic relations, which are braided extensions of
the Yang-Baxter equations. The relationship of QIS with stat mech and CFT
gives practical results in finding finite-temperature and finite size
corrections of the quantum model as well as its
 underlying conformal properties.  
 
Recently found  link of the QIS with diverse subjects like link and knot
polynomials \c{wadatipr}, 
reaction-diffusion processes \c{reacrit}, Seiberg-Witten model \c{sw}
 etc. are also 
becoming more and more important.

 \end{document}